\documentclass[twocolumn,showpacs,preprintnumbers,amsmath,amssymb]{revtex4}
\usepackage{graphicx}
\usepackage{bm}
\usepackage{latexsym}
\usepackage{epsfig}
\usepackage{natbib}

\newcommand{\CC}{\Lambda}

\begin{document}

\providecommand{\U}[1]{\protect\rule{.1in}{.1in}}
\newcommand{\be}{\begin{equation}}
\newcommand{\ee}{\end{equation}}
\newcommand{\OM}{\Omega_M}
\newcommand{\Omm}{\Omega_m}
\newcommand{\Omo}{\Omega_m^0}
\newcommand{\OL}{\Omega_{\Lambda}}
\newcommand{\OLo}{\Omega_{\Lambda}^0}
\newcommand{\rc}{\rho_c}
\newcommand{\rco}{\rho_{c}^0}
\newcommand{\rmo}{\rho_{m0}}
\newcommand{\rmm}{\rho_{m}}
\newcommand{\mincir}{\raise
-3.truept\hbox{\rlap{\hbox{$\sim$}}\raise4.truept\hbox{$<$}\ }}
\newcommand{\magcir}{\raise
-3.truept\hbox{\rlap{\hbox{$\sim$}}\raise4.truept\hbox{$>$}\ }}
\newcommand{\newtext}[1]{\text{#1}}
\newcommand{\newnewtext}[1]{\text{#1}}
\newcommand{\newnewnewtext}[1]{\text{#1}}
\newcommand{\newfinal}[1]{\text{#1}}
\newcommand{\rM}{\rho_m}
\newcommand{\pM}{P_m}
\newcommand{\pL}{P_{\CC}}


\title{Cosmic acceleration without dark energy: Background tests and thermodynamic analysis}

\author{J. A. S. Lima} \email{jas.lima@iag.usp.br}
\affiliation{Departamento de Astronomia, Universidade de S\~ao Paulo, 55080-
900, S\~ao Paulo, SP, Brazil}

\author{L. L. Graef} \email{leilagraef@usp.br}
\affiliation{Instituto de F\'{i}sica, Universidade de S\~ao Paulo, Rua do Mat\~ao travessa R, 05508-090, S\~ao Paulo, SP, Brazil}

\author{D. Pav\'on} \email{diego.pavon@uab.es}
\affiliation{Departamento de F\'{\i}sica, Universidad Aut\'{o}noma
de Barcelona, 08193 Bellaterra (Barcelona), Spain}

\author{Spyros Basilakos} \email{svasil@academyofathens.gr}
\affiliation{Academy of Athens, Research Center for Astronomy and Applied
Mathematics, Soranou Efesiou 4, 11527, Athens, Greece}

\begin{abstract}
A new cosmic scenario with gravitationally induced particle
creation is proposed. In this model the Universe evolves from an
early to a late time de Sitter era, with the recent  accelerating
phase driven only by the negative creation pressure associated
with the cold dark matter component. The model can be interpreted
as an attempt to reduce the so-called cosmic sector (dark matter
plus dark energy) and relate the two cosmic accelerating phases
(early and late time de Sitter expansions).  A detailed
thermodynamic analysis including possible quantum corrections is
also carried out. For a very wide range of the free parameters, it
is found that the model presents the expected behavior of an
ordinary macroscopic system in the sense that it approaches
thermodynamic equilibrium in the long run (i.e., as it nears the
second de Sitter phase). Moreover, an upper bound is found for the
Gibbons-Hawking temperature of the primordial de Sitter phase.
Finally, when confronted with the recent observational data, the
current `quasi'-de Sitter era, as predicted by the model, is seen
to pass  very comfortably the cosmic background tests.
\end{abstract}
\maketitle

\section{Introduction}
In the standard cosmological model, the universe is homogeneous
and isotropic  and  the main sources of the gravitational field is
a mixture of ideal fluids containing a baryonic plus cold dark
matter (CDM) components, and a cosmological constant, $\Lambda$.
Due to Lorentz's invariance, the latter is endowed with a negative
pressure that accounts for the present state of accelerated
expansion \cite{review}. With just an additional free parameter,
$\Lambda$,  the cosmic concordance lambda cold dark matter model
($\Lambda$CDM) fits rather well the current astronomical data from
supernovae type Ia, baryon acoustic oscillations (BAO), cosmic
microwave background (CMB), galaxy clusters evolution, and
complementary observations
\cite{supernovae,Suzuki:2011hu,CMB,Ade13,clusters,CO1,CO2}.

Nevertheless, there are severe and profound drawbacks related to a
finite but incredibly small value  of $\Lambda$. Firstly, attempts
to associate it to the vacuum energy density estimated by quantum
field theory leads to a discrepancy of $50$ to $120$ orders of
magnitude with respect to its observed value, about $3 \times
10^{-11}$eV$^{4}$. This implies an extreme fine-tuning problem
giving rise to the so-called cosmological constant problem, which
also requires an improbable cancellation by some unknown physical
mechanism \cite{finetunning}.  This is why generic proposals
replacing $\Lambda$ by some evolving field termed ``dark energy"
were suggested by many authors, however, the true nature of this
field still remains elusive \cite{reviews}. Secondly,  there is
also the coincidence problem which is related to the question of
``why are the energy densities of pressureless matter, $\rho_{m}$,
and vacuum, $\rho_{\Lambda} = \Lambda/8\pi G$, of the same order
precisely today in spite of the fact that they evolve so
differently with expansion?" \cite{coincidence}. Proposals to
alleviate  such problems include decaying vacuum models which
promote the cosmological constant to a field, $\Lambda(t)$, that
varies with time in a suitable manner, and many interacting scalar
field descriptions of dark energy
\cite{OT86,CLW92,Waga93,LM1,OverduinCooper98,ShapSol00,WM,BPS09,vd2,vd3,stab2,graef},
as well as a single fluid with an antifriction dynamics
\cite{prd-2001}.

On the other hand, the recent accelerating phase of the Universe
was probably not the only one. According to the standard
cosmological model, the Universe must have experienced a very
brief period $(\sim 10^{-30}s)$ of fast accelerated expansion
shortly after the big bang, responsible for  the observed
homogeneity and isotropy of the Universe on large scales, its
spatial flatness, and the spatial fluctuations in temperature of
the cosmic background  (CMB) radiation. Hence, the need arises for
introducing a further unknown energy component to account for
this. Unfortunately this ``solution'' comes about with several new
problems, like the initial conditions, the graceful exit and
multiverse problems,  a combination leading to the existence of
new fine tunings and some conceptual problems \cite{ISL}.

Because of  such difficulties, here we suggest another well known
proposal for the cosmic acceleration, the gravitational particle
production mechanism. The microscopic description of this process
was pioneered by  Schrodinger's \cite{sch}, and developed by
Parker and others based on quantum field theory in curved
spacetimes \cite{parker1,parker2,parker3,parker}. There is still
an open question concerning  the actual range of the particle
creation effects in cosmology, since such process involves
non-equilibrium quantum field theory in curved space-times, which
was not, by now, theoretically developed. We know, however that
such a mechanism must in fact take place, and therefore, its
possible effects on cosmology should be considered, at least from
a phenomenological point of view \cite{graef2}.

A macroscopic description of  the particle production mechanism by
the gravitational field was also discussed long ago by Prigogine
and collaborators \cite{prigogine}. Later on, Calv\~ao, Lima and
Waga proposed a covariant description \cite{LimaCW92}, and the
physical difference between particle production and the bulk
viscosity mechanism was clarified by Lima and Germano \cite{LG92}.
The particle production process is classically described by  a
back reaction term in the Einstein field equations whose negative
pressure may provide a self-sustained mechanism of cosmic
acceleration. Indeed, since the middle of the nineties, many
phenomenological accelerating scenarios have been proposed  in the
literature \cite{AL99}.

Some years ago, it was shown that phenomenological particle
production  can explain not only  the present era of cosmic
acceleration but also  provide a viable alternative to the
concordance $\Lambda$CDM model \cite{LSS,LJO}. Recently, it has been argued that
CCDM is observationally degenerate with respect to $\Lambda$CDM (dark degeneracy) even at a perturbative level \cite{waga2014}. 
In principle, one may also think that the mechanism of particle production could
account  not only for the late time cosmic acceleration but also
for the primeval one (i.e., early inflation). In this case,
besides evading the problems related to the cosmological constant
it may also have some advantages with respect to  standard
inflation. In fact, as shown by Lima, Basilakos and Costa
\cite{LBC12} (LBC hereafter) gravitationally-induced  particle
production in the course of  the expansion can be responsible by
the instability of the initial de Sitter state which evolves
smoothly to the standard radiation phase when the production of
massless particles is suppressed. More interesting, not only the
horizon problem is solved, but also the production of relativistic
particles during inflation  avoids the supercooling and the need
for a subsequent reheating phase, thereby solving in a natural way
the exit problem. Qualitatively, such a scenario resembles a
variant of the so-called ``warm inflation" \cite{Berera}.

The new  cosmological scenario proposed here  generalizes the LBC
model whose thermodynamic behavior was investigated by Mimoso and
Pav\'on \cite{MP13}. As  with the LBC cosmology, this novel
scenario is also complete in the sense that  it describes the
cosmic evolution from an early to a late time de Sitter phase with
the  accelerating stages powered by the creation of particles by
the gravitational field. As we shall see, the Universe starts from
a nonsingular unstable de Sitter phase thereby being free of the
horizon problem, and, smoothly evolves to the standard
radiation-dominated phase. As the Universe keeps on expanding, the
radiation component becomes subdominant and pressureless dark
matter takes over. It is then when the production of  CDM
particles gets triggered. Finally, the Universe approaches  the
second de Sitter era of expansion characterized by thermodynamic
equilibrium.

The generalization of the complete particle production scenario
intends to improve this model by providing a more embracing and
dynamical cosmic evolution, maintaining, however, the ability of
recovering the $\Lambda$CDM dynamics for certain values of the
free parameters. Especially, at the early Universe,  we expect
that the proposed generalization will provide a way toward the
conciliation of the model with  recent CMB observations by the
Planck satellite \cite{Ade13,ISL}.


In order to test the range of the parameter space of the model we
also perform a thermodynamic analysis based on the  generalized
second law (GSL) of thermodynamics. This law, first formulated for
black holes and their environment \cite{jakob} and later extended
to cosmic horizons \cite{gslcosmic}, establishes that the entropy
of the system plus that of the causal horizon enveloping it should
never decrease. Further, in the last stages of the evolution the
total entropy should also be a concave function. Otherwise, the
total entropy (system plus horizon) would increase unbounded
without ever reaching equilibrium -the state of maximum entropy
compatible with the constraints upon the system \cite{grgnd}. Our
aim is to explore which restrictions (if any) the GSL plus the
concavity requirement impose on the free parameters of the
cosmological model.

The paper is organized as follows. Next section briefly introduces
the basics of phenomenological particle production in cosmology.
Section III considers this effect in  the early Universe after the
initial de Sitter expansion. Section IV studies the corresponding
constraints imposed by the second law of thermodynamics. Section V
generalizes the scenario of Ref. \cite{LBC12}. The thermodynamic
analysis based on the GSL is carried out with emphasis on the
transition from  matter dominated  to the second de Sitter phase.
Based on recent observations we also determine some constraints on
the  main parameters of the model. Finally, section VI summarizes
our findings.

\section{Cosmic dynamics in models with particle production}
Let us consider a flat, homogeneous and isotropic
Friedmann-Robertson-Walker (FRW) universe, whose matter content is
endowed with the mechanism of particle production.
 In this case the Friedmann equations can be written as \cite{LimaCW92,LG92}:
\begin{equation}
8\pi G{\rho} = 3\frac{\dot{a}^2}{a^2},
\label{friedmann1}
\end{equation}
\begin{equation}\label{pressure}
8\pi G({p} + p_{c}) = -2\frac{\ddot{a}}{a} - \frac{\dot{a}^2}{a^2},
\end{equation}
where ${\rho}$ and $p$ are the energy density and the equilibrium
hydrostatic pressure, $a$ is the cosmic scale factor (the over-dot
means derivative with respect to cosmic time) and $p_{c}$ is the
creation pressure which is related to the gravitationally induced
process of particle production.

As a consequence, the energy conservation law generalizes to
\begin{equation}
\dot{\rho}+3H(\rho+p+p_{c})=0 \, .
\label{conservation}
\end{equation}

Recently, a great deal of attention has been paid to scenarios
driven by `adiabatic' particle production. In this case, particles and
entropy are generated  but the  entropy per particle does not
vary. Under such `adiabatic condition', the creation pressure can be
written as \cite{LSS,LBC12,MP13}
\begin{equation}
p_{c}=- (\rho+p)\, \frac{\Gamma}{3H}\, ,
\end{equation}
where the positive-definite quantity $\Gamma$ denotes the rate of
particle production. We shall denote the latter by $\Gamma_{r}$
and $\Gamma_{m}$ during the early and late phases of expansion,
respectively.

By assuming the usual equation of state, $p= w \rho$, it is
readily checked from  Eqs. (1)-(2) and (4) that  the evolution of the
Hubble parameter is governed by the differential equation

\begin{equation}
\dot{H} + \frac{3(1+w)}{2}H^{2} \left(1 -
\frac{\Gamma}{3H}\right)=0.
\end{equation}
Thus, this scenario will be fully determined once $w$ and $\Gamma$
are specified. Note that for $w = $ constant and $\Gamma \ll 3H$ the
standard evolution, $a(t) \propto t^{\frac{2}{3(1+ w)}}$,
is readily recovered.

\section{Production of particles in the early universe}
In the previous section we have seen that  the ratio $\Gamma/3H$
is the key to determine the dynamics of the model. Due to the
absence of a rigorous quantum field theory in curved space-time,
including (non-equilibrium) back reaction, from which the
expression of $\Gamma$ should be calculated, in \cite{LBC12} a
phenomenological early universe model with a particle creation
rate, given by
\begin{equation}
\frac{\Gamma_{r}}{3H}=\frac{H}{H_{I}}\, ,
\end{equation}
was introduced. Here,  $H_{I}$ is the constant inflationary
expansion rate associated to the initial de Sitter phase ($H \leq
H_{I}$), and $\Gamma_{r}$ is the creation rate of relativistic
particles in the transition from the early de Sitter stage to the
radiation dominated phase.

The ratio $H/H_{I}$ takes into account that the particle
production must be strongly suppressed ($\Gamma_{r}/3H \ll 1$)
when the Universe enters  the radiation phase. This is so because,
according to Parker's theorem, massless particles cannot be
quantum-mechanically produced in that phase \cite{parker}.

This photon creation model at early times  provides an interesting
description of the Universe evolution \cite{LBC12}. The latter
starts with a de Sitter expansion without initial singularity.
This expansion, due to the creation of massless particles, becomes
unstable and  the universe subsequently enters the conventional
radiation-dominated era.

At this point, it is interesting to consider  the following
generalization of (6)
\begin{equation}
\frac{\Gamma_{r}}{3H}=\left(\frac{H}{H_{I}}\right)^{n},
\end{equation}
where $n$  is a nonnegative constant parameter to be constrained
by observational data.  The above expression reduces to the
original model \cite{LBC12} in the particular case of $n=1$.

Replacing this more general expression for the creation rate  in
Eq. (5), it follows that the evolution of the Hubble parameter in
this case ($w = 1/3$) is governed by
\begin{equation} \label{dotH}
\dot{H} + 2H^{2} \left[1 -
\left(\frac{H}{H_{I}}\right)^{n}\right]= 0.
\end{equation}
As expected, for $H=H_{I}$ we obtain an unstable de Sitter
solution, $\dot{H}=0$. The creation of particles, as before,
causes a dynamic instability that leads to a transition from a de
Sitter regime to the radiation dominated era.

By integrating the last equation we obtain
\begin{equation}
H=\frac{H_{I}}{(1+Da^{2n})^{1/n}}, \label{H(a)}
\end{equation}
or, alternatively,
\begin{equation}\label{HS1b}
\int_{a_\star}^a\frac{d\tilde{a}}{\tilde{a}}\left[1+D\,\tilde{a}^{2\,n\,}\right]^{1/n}= H_It
\end{equation}
where $D$ is a positive definite constant, $t$ is the time elapsed
after the end of the inflationary era, denoted by $t_\star$, hence
$a_\star=a(t_\star)$. Using the condition $H(a_\star)\equiv
H_\star$, the constant of integration $D$ is found to be

\begin{equation}\label{eq:defD}
D=a_\star^{-2\,n\,}\left[\left(\frac{H_I}{H_\star}\right)^n-1\right]\,.
\end{equation}
Integration of  Eq.(\ref{HS1b}) yields the cosmic time $t(a)$ in terms of the scale factor \cite{Transform}
\begin{equation}\label{tn}
t(a)= \frac{\left(1+D\,a^{2n}\right)^{\frac{1+n}{n}}}{2 H_I\,D\,
a^{2n}} \times
F\left[1\,,1\,,1-\frac{1}{n}\,,\frac{-1}{D\,a^{2n}}\right]\,,
\end{equation}
where $F[\alpha_1,\alpha_2,\alpha_3,z]$ is the Gauss
hypergeometric function.

It is easy to show that the correct transition from an early de
Sitter to the radiation phase is obtained for any positive value
of $n$. Thus, our analysis shows that this generalized model can
describe the dynamics of the early universe free of the big bang
singularity due to $D\ne 0$, and that it overcomes the horizon
problem.

Specifically, the Universe starts from an unstable inflationary
phase [$H\simeq H_{I}$ with $a(t)\simeq a_{\star}  e^{H_It}$]
powered by the huge value $H_I$ which may connected to the scale
of a grand unified theory or even the Planck scale, then it
deflates (with a massive production of relativistic particles),
and subsequently evolves towards the radiation-dominated  era,
$a\sim t^{1/2}$ (i.e., $H\simeq a^{-2}$), for $Da^{2n}\gg 1$ in
Eq.(\ref{H(a)}). Hence,  there is  ``graceful exit'' from the
inflationary stage.

\section{Thermodynamic Analysis of the Early Universe}
Let us now discuss the thermodynamic behavior of the model in the
radiation era. Thermodynamics tells us that the entropy of
isolated systems can never diminishes, and it is concave, at least
during the last stage of approaching equilibrium (otherwise no
entropy maximum could ever be achieved).

Recently it was demonstrated that cosmological apparent horizons
are also endowed with thermodynamical properties \cite{GSL}. One
can relate a temperature and entropy to the apparent horizon
analogous to the ones  associated to the black hole event horizon.
Unlike the event horizon, the cosmic apparent horizon always
exists and it coincides with the event horizon in the case of a
last de Sitter space. So,  in accordance with the GSL,  the total
entropy $S$ must include the entropy of all sources, that is, the
fluid inside the apparent horizon and the entropy of the apparent
horizon itself. Denoting by  $S_{\gamma}$ the entropy when the
Universe is radiation-dominated and $S_h$ the apparent horizon
entropy, it  thus follows that $S = S_{\gamma}\, + \, S_{h}$.

The radiation phase is followed by a matter dominated era that
eventually will transit to a second de Sitter phase. Accordingly,
we expect that in the radiation phase the entropy increases and be
a convex function of the scale factor, i.e., $ S'
>0$ and $S'' > 0$ (a prime means $d/da$). Were it concave,
the Universe would have attained a state of thermodynamic
equilibrium (maximum entropy) and would stayed in it for ever
unless forced by some ``external agent". However, as is
well-known, during the radiation phase the production of particles
is suppressed \cite{parker}; so in this model there would be no
external agent to remove the system from thermodynamic
equilibrium. This is why we expect the entropy to be convex in
this phase.

The entropy of the apparent horizon is given by $S_{h}=k_{B}\,
{\cal A}/4 \ell_{pl}^2$  \cite{bak-rey}, where ${\cal A} =4\pi
r_{h}^{2}$ is the area of the horizon, $k_{B}$ the Boltzmann's
constant, $\ell_{pl}$ the Planck's length, and $r_{h}$ the radius
of the horizon. In our case, a spatially-flat Universe, the latter
coincides with the Hubble horizon, $H^{-1}$.

On the other hand, the entropy of the radiation fluid can be
obtained from Gibbs's equation,
\begin{equation}
T_{\gamma}dS_{\gamma}=d(\rho_{\gamma} V)\, + \, p_{\gamma}\, dV,
\label{gibbs}
\end{equation}
where $V=4\pi/(3H^{3})$ is the spatial volume enclosed by the
horizon, $T_{\gamma}$ the radiation temperature, $p_{\gamma} =
\rho_{\gamma}/3$ with
\begin{equation}
\rho_{\gamma} = \frac{\rho_{I}}{(1\, + \, Da^{2n})^{2/n}} \, ,
\quad {\rm and} \quad \rho_{I} \equiv \frac{3H_{I}^{2}}{8 \pi G}.
\label{rho-gamma}
\end{equation}
In arriving at this expression  use of Eqs. (\ref{friedmann1}) and
(\ref{H(a)}) was made.

By deriving $S_{h}$  and using Eq. (\ref{H(a)}) again, we obtain
\begin{equation}\label{sprime1h1}
S'_{h} = \frac{4 k_{B} \pi}{\ell^{2}_{pl} H^{2}_{I}}\, D \,
a^{2n-1}(1+Da^{2n})^{\frac{2}{n} - 1}.
\end{equation}
Clearly, $S'_{h} >0$ regardless the value of $n$.

On its part the radiation temperature obeys,
\begin{equation}\label{Tgamma}
T_{\gamma} = \frac{T_{I}}{(1\, + \, D a^{2n})^{\frac{1}{2n}}}
\end{equation}
where   $T_{I}$ is the initial temperature in the de Sitter phase.
Note that for $n=1$ the LBC expression is recovered (see  Eq. (13)
there). Obviously, for $Da^{2n} \gg 1$ (well inside the radiation
era) we recover the standard radiation result, i.e., $T_{\gamma}
\propto a^{-1}$.

From Gibbs's equation (\ref{gibbs}) it follows that
\begin{equation}\label{sgammaprime}
T_{\gamma}\, S'_{\gamma} = \frac{16 \pi}{3} \rho_{I}
\frac{D}{H_{I}^{3}} \, a^{2n-1}\, (1 \, + \, D
a^{2n})^{\frac{1}{n}-1} \, ,
\end{equation}
that is to say, $S'_{\gamma} >0 $ irrespective of the value of
$n$.

To discern whether $n$ gets constrained by the convexity of the
total entropy we must determine the sign of the second derivatives
of both entropies. From (\ref{sprime1h1}) we readily get
\begin{equation}
S''_{h} = C \, D\, a^{2(n-1)} (1\, + \, D a^{2n})^{\frac{2}{n}-2}
\, \left[ 3 D \,  a^{2n}\, + \, 2n-1 \right] \, ,
\label{sprimeprime2h}
\end{equation}
and from (\ref{Tgamma}) and (\ref{sgammaprime})
\begin{equation}
S''_{\gamma} = \frac{16\pi}{3}\, \frac{\rho_{I} D}{T_{I}
H_{I}^{3}}\, a^{2(n-1)}\, (1+Da^{2n})^{\frac{3}{2n}-1} \left[
\frac{2(Da^{2n}+n)\, - \, 1}{1\, + \, Da^{2n}}\right].
\label{sgammaprime2}
\end{equation}
Thus the positivity of both second derivatives is ensured whenever
$\, n>1/2$.

Altogether, while the GSL does not set any constraint on $n$ the
convexity of the total entropy during the radiation era do poses a
lower bound on this parameter.

\subsection{Quantum corrections}
It is well-known that quantum effects generalize the
Bekenstein-Hawking entropy law for black holes to the expression
\begin{equation}\label{qcorrections}
S_{h}=k_{B}\left[\frac{{\cal A}}{4 \ell_{pl ^{2}}} - \frac{1}{2}
{\rm ln} \left(\frac{{\cal A}}{\ell_{pl} ^{2}}\right)\right],
\end{equation}
plus higher order terms \cite{QC,QC2}. As pointed out in
\cite{MP13} the same should apply to causal cosmic horizons. Here
we analyze whether our results remain valid when such corrections
are not overlooked.

A simple calculation in the context of our scenario yields
\begin{equation} \label{sprimehqc}
{S'_{h}}=\frac{k_{B}\pi}{{\ell_{pl}}^{2}}\frac{4}{a \, H^{2}}
\left(1 - \left(\frac{H}{H_{I}}\right)^{n}\right) \left[1 -
\frac{\ell_{pl}^{2} H_{I}^{2}}{2\pi(1+Da^{2n})^{2/n}}\right].
\end{equation}
It is immediately seen that the presence of the factor
$\ell_{pl}^{2}$ in the numerator of the second term in the square
parenthesis renders the said term negligible. Thereby our approach
is robust against quantum modifications to the horizon entropy in
the early Universe.

Moreover, in the limit $a \rightarrow 0$ the condition $S'_{h} >0$
implies the upper bound on the initial expansion rate,
$H_{I}<\sqrt{2\pi}/\ell_{pl}$, independent on $n$. Thus, the
generalization of the model does not alter the original sensible
result obtained in \cite{MP13}: The  initial Hubble factor cannot
be arbitrarily large; his squared value is limited by Planck's
curvature.

It is also interesting that the temperature $T_I$ appearing in the
expression (\ref{Tgamma}) has also a natural upper limit imposed
by the quantum corrections discussed here. In fact, recalling that
the initial temperature of the Universe in our scenario can be
associated to the Gibbons-Hawking result \cite{gslcosmic},  $T_I =
H_I/2\pi$ (see LBC), it is easy to check from the above inequality
that $T_{I}<1/\sqrt{2\pi}\ell_{pl}$. In other words, the quantum
corrections to the usual entropy formula imply that the initial
temperature of the Universe in our model is slightly smaller than
Planck's temperature, as it should be expected from a classical
description.


\section{A complete generalized cosmological scenario}
In the context of particle production models it was shown in the
LBC paper that the two eras of accelerated expansion  can be
closely related through a single expression for the particle
creation rate. The model was called ``a complete cosmological
scenario without dark energy" \cite{LBC12}. The resulting
cosmology showed consistency with the observational  data both at
the background and perturbative level. The phenomenological
expression proposed in that work for the particle creation rate
was
\begin{equation}
\frac{\Gamma}{3H}=\frac{H}{H_{I}} + \left(\frac{H_{f}}{H}\right)^{2}.
\end{equation}
The first and second terms on the right hand side dominate at
early and late times, respectively. Since we have already shown
that the proposed extension of the first term is compatible with
GSL, we now take a step further by generalizing  the second term
as well.

Let us now  introduce  a new extended  scenario in which the
expression for the particle production rate also encompasses the
whole whole cosmic expansion, namely:
\begin{equation} \label{complete}
\frac{\Gamma}{3H}=\left(\frac{H}{H_{I}}\right)^{n} + \left(\frac{H_{f}}{H}\right)^{m},
\end{equation}
where the free parameter $m$ must be non-negative to lead to an
acceptable matter-vacuum expansion history.  However,  the above
(\ref{complete}) encompasses several dynamical possibilities
involving decaying vacuum models   $\Lambda(t)$ (i.e., with
$\dot{\Lambda}(t) < 0$), which also solves (or, at least,
alleviate) the coincidence problem. In fact, as recently shown in
\cite{graef2}, particle production by the gravitational field can
also mimic the dynamics of $\Lambda(t)$ models by the
correspondence
\begin{equation}
\frac{\Lambda(t)}{3H^{2}}= \frac{\Gamma}{3H},
\end{equation}
where the evolution of $\Lambda(t)$  is usually described by a
phenomenological power law in $H$.  Two previous examples
considered in the literature are, $\Lambda(t)=\beta H^{2}$
\cite{CLW92}, and $\Lambda(t)= \beta H$ \cite{saulo}. In both
cases $\beta$ is a positive constant. Therefore, by allowing
$\Gamma_{m}/3H$ to be described at late times by a general power
law in $H$, we can also include in the description some physically
viable decaying vacuum dynamics (see also the discussion in
\cite{BPS09}). Such models have been proposed in order to
alleviate the coincidence and $\Lambda$ problems, and,
simultaneously,  to describe the cosmic expansion closely to the
$\Lambda$CDM  at recent times.

At late times, pressureless matter takes over radiation as the
dominant energy component (we denote the former by a subscript
$m$), and the Hubble factor satisfies $ H_{f} \leq H \ll H_{I}$.
Accordingly, at the epoch when the second term in (\ref{complete})
dominates the evolution of $\, H \, $ is dictated by
\begin{equation} \label{evol}
\dot{H}=-\frac{3}{2}H^{2}\left[1-\left(\frac{H_{f}}{H}\right)^{m}\right].
\end{equation}
Its solution in terms of the scale factor can be written as
\begin{equation}\label{H(a)late}
H(a) = \left(C \, a^{-3m/2}\, + \, H_{f}^{m}\right)^{1/m}\, ,
\end{equation}
where $C= H^{m}_{0}\,- \, H^{m}_{f}$.

By evaluating  the above expression at the present time [$a =
a_{0} = 1$, $H(a=1)=H_{0}$] we obtain
\begin{equation}\label{H1}
H(a) = H_{0}\left(\Omega_{m0}a^{-3m/2}\, + \, {\tilde\Omega_{\Lambda 0}}\right)^{1/m}\, ,
\end{equation}
where  $\Omega_{m0}= 1\, -\, (H_{f}/H_{0})^{m}$ and
${\tilde\Omega_{\Lambda 0}}=1-\Omega_{m0}$. Obviously, for $m=2$,
we recover, in an effective way, the evolution of the $\Lambda$CDM
model. By using the above Hubble parameter,  we have performed a
joint statistical analysis involving the latest observational
data, namely:  SNIa-Union2.1 \cite{Suzuki:2011hu}, BAO
\cite{Blake:2011en,Perc10} and the Planck CMB shift parameter
\cite{Ade13,Shaf13}. The corresponding covariances can be found in
Basilakos {\it et al.} \cite{BasNes13} for the SNIa/BAO data and
in \cite{Shaf13} for the Planck CMB shift parameter, respectively.
 Specifically, the joint $\chi^2_{t}$ function is given by
$\chi^{2}_{t}({\bf
p})=\chi^{2}_{SNIa}+\chi^{2}_{BAO}+\chi^{2}_{CMB}$, where ${\bf
p}$ is the statistical vector that contains the free parameters of
the model, namely ${\bf p}=(\Omega_{m0},m)$. The expressions of
the individual chi-square functions: $\chi^{2}_{SNIa}$,
$\chi^{2}_{BAO}$ and $\chi^{2}_{CMB}$ can be found in
\cite{BasNes13}.

As it turns out, the overall likelihood function peaks at
$\Omega_{m0}=0.283\pm 0.012$, $m=1.934 \pm 0.06$ with
$\chi_{t,min}^{2}(\Omega_{m0},m)\simeq 563.6$, resulting in a
reduced value of $\chi^{2}_{t,min}/dof\sim 0.96$. Figure 1 shows
the 1$\sigma$, 2$\sigma$ and $3\sigma$ confidence contours in the
$(\Omega_{m0},m)$ plane of the joint analysis. Alternatively,
considering the $\Lambda$CDM theoretical value of $m=2$ and
minimizing with respect to $\Omega_{m0}$ we find
$\Omega_{m0}=0.292\pm 0.008$ (see the inset in Fig. 1) with
$\chi^{2}_{t,min}(\Omega_{m0})/dof \simeq 567.5/585$.

We also made use of, the relevant to our case, {\em corrected}
Akaike information criterion (AIC) \cite{Akaike1974}, defined, for
the case of Gaussian errors, by
\begin{equation}
{\rm AIC}=\chi^2_{t,min}+2 k \, ,
\end{equation}
where $k$ denotes the number of free parameters. A smaller value
of AIC indicates a better model-data fit. However, small
differences in AIC are not necessarily significant and therefore,
in order to assess, the effectiveness of the different models in
reproducing the data, one has to investigate the model pair
difference $\Delta$AIC$ = {\rm AIC}_{y} - {\rm AIC}_{x}$. The
higher the value of $|\Delta{\rm AIC}|$, the higher the evidence
against the model with higher value of ${\rm AIC}$; a difference
$|\Delta$AIC$| \ge 2$ indicates a positive such evidence while if
$ |\Delta$AIC$| \ge 6$ the evidence is strong. In its turn,  if $
|\Delta$AIC$| \le 2$ consistency among the two models under
comparison should be assumed. In our case, the value of the
particle creation model AIC$_{\Gamma}$($\sim 567.6$) is smaller
than the corresponding one AIC$_{\Lambda}$($\sim 569.5$), which
suggests that the particle creation cosmological model provides a
better fit to the current data than the concordance $\Lambda$CDM
cosmology. On the other hand, the $|\Delta {\rm AIC}|$=$|{\rm
AIC}_{\Lambda}-{\rm AIC}_{\Gamma}|\simeq 2$ value points out that
the said data can hardly tell apart  the $\Lambda$CDM from the
particle creation model.

\begin{center}
\begin{figure}
\mbox{\epsfxsize=10.2cm \epsffile{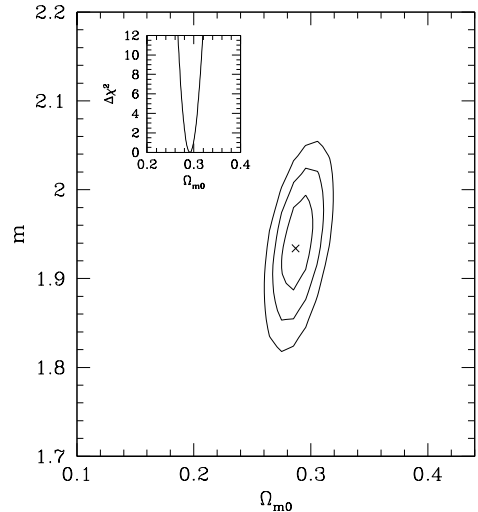}}
\caption{Likelihood
contours for $\Delta \chi^2=\chi^{2}-\chi^{2}_{t,min}$ equal to
2.32, 6.18 and 11.83, corresponding to 1$\sigma$, 2$\sigma$ and
$3\sigma$ confidence levels, in the $(\Omega_{m0},m)$ plane using
the SNIa/BAO/CMB$_{shift}$ overall statistical analysis (see
text). The cross signals the best fit model
$(\Omega_{m0},m)=(0.283,1.934)$. The inset shows the solution
space for the concordance $\Lambda$CDM model.}
\end{figure}
\end{center}

Finally, in order to see which values of the parameter $m$ are
thermodynamically allowed we write the first derivative of the
entropies of the horizon and matter. The first one is,
\begin{equation}\label{sprimehlate}
S'_{h}= \frac{2k_{B}\pi}{\ell_{pl}^{2}a H^{2}}\left[\frac{3}{2}
\left(1-\left(\frac{H_{f}}{H}\right)^{m}\right)\right]\, .
\end{equation}
As for the entropy of the matter fluid  inside the horizon, it
suffices to realize that every single dust particle contributes a
given bit, say, $k_{B}$, \cite{MP13}. So, $S_{m}=k_{B}4\pi
r_{h}^{3}n_{p}/3$, where the number density of particles, $n_{p}$,
obeys the conservation equation $n'_{p} = (n_{p}/a)[(\Gamma_{m}/H)
\, - \, 3]$ with $\Gamma_{m} = H (H_{f}/H)^{m}$. Hence,
\begin{equation}\label{sprimemlate}
S'_{m} = \frac{4k_{B}\, \pi n_{p}}{3 a H^{2}}\left[\frac{3}{2}
\left(1-\left(\frac{H_{f}}{H}\right)^{m}\right)\right] \, .
\end{equation}
As in the radiation  case, the  GSL, $ S' = S'_{m} \, + \, S'_{h}
\geq 0$, only constraints $m$ to be positive (something previously
demanded to satisfy  cosmic dynamics). Now, the condition that the
total entropy approaches a maximum in the long run  does not
impose any further condition on $m$. Indeed, from Eq.
(\ref{H(a)late}) we see that $H \rightarrow H_{f}$ when $a
\rightarrow \infty$, therefore both $S'_{h}$ and $S'_{m}$ tend to
zero in that limit. On the other hand, since both first
derivatives are positive for finite scale factor, we conclude that
$S'$ tends to zero from below; hence, $S''(a\rightarrow \infty)
\leq 0$ which can be realized for positive values of $m$ only.
Altogether, the generalized complete model \cite{LBC12} is
consistent with thermodynamics also at late times for any positive
value of the free parameter $m$.

\section{Conclusions}
In this work  a generalized complete cosmological scenario of
particle production, evolving from de Sitter to de Sitter was
presented. Its thermodynamic viability, according to the GSL and
the thermodynamic requirement that the entropy of the total system
(fluid plus horizon) tends to a maximum in the long run, was
investigated. As it turns out,  the parameter $n$
[see Eq.(\ref{complete})] must be larger than $1/2$ while any positive
value of $m$ shows compatibility with thermodynamics. Further, the
inclusion of quantum corrections \cite{QC,QC2} in the limit $a
\rightarrow 0$ sets a very reasonable  upper bound on the initial
Hubble rate, $H_{I}$, and on the Gibbons-Hawking temperature,
$T_{I}$, which cannot be obtained by purely classical methods.

The statistical analysis of the model shows that, when confronted
with current observational data, it performs not less well than
the concordance $\Lambda$CDM model. We believe that the cosmological
model proposed here provides a viable and complete scenario in the
sense that it closely relates the two accelerated phases of the
Universe through a single and simple {\em ansatz}, Eq.
(\ref{complete}). Moreover, in simplifying the dark sector it
evades the coincidence and the cosmological constant problems of
the $\Lambda$CDM model, and the need to introduce unknown
components, like dark energy, or unobserved extra dimensions.

\begin{acknowledgments}
J.A.S.L. is partially supported by CNPq and
FAPESP under grants 304792/2003-9 and 04/13668-0, respectively, and L.L.G.
by a PhD fellowship from FAPESP (grant 09380-8).
D.P. acknowledges support from the ``Ministerio de Econom\'{\i}a y
Competitividad, Direcci\'{o}n General de Investigaci\'{o}n
Cient\'{\i}fica y T\'{e}cnica", Grant N$_{0.}$ FIS2012-32099.
SB acknowledges support by the Research Center for
Astronomy of the Academy of Athens
in the context of the program  ``{\it Tracing the Cosmic Acceleration}''.
\end{acknowledgments}

\vspace{0.5cm} \centerline{\bf APPENDIX} \vspace{0.5cm}
\centerline{\bf Scalar field description in the Early Universe}
\vspace{0.3cm}
From the section II it became clear that the particle creation
model is capable to overcome the basic cosmological problems.
Traditionally, it is useful to represent the cosmic evolution in a
field theoretical language, i.e., in terms of the dynamics of an
effective scalar field ($\phi$). In a point of fact, all the
dynamical stages discussed here can be described through a simple
scalar field model. For a similar analysis in the case of bulk
viscosity see \cite{ZIM00}, and for the equivalent decaying
$\Lambda(t)$-models see Refs. \cite{vd3,MaiaLima02}.

To begin with, let us replace $\rho$ and $p_{tot}=p+p_{c}$ in Eqs.
(\ref{friedmann1}) and (\ref{pressure}) by the corresponding
scalar field expressions
\begin{equation}
\label{scal} \rho \rightarrow \rho_{\phi} =
\frac{\dot{\phi}^{2}}{2} + V(\phi), \;\;\;\;\;\; p_{tot}
\rightarrow p_{\phi} =\frac{\dot{\phi}^{2}}{2} - V(\phi) \;.
\end{equation}
Substituting, the above into the Friedmann's equations we can
separate the scalar field contributions and express them in terms
of $H$ and $\dot{H}$, i.e.,
\begin{equation}
\dot{\phi}^{2} =-\frac{1}{4\pi G}\dot{H} \;, \label{ff3}
\end{equation}
\\ \
\begin{equation}
\label{Vz} V=\frac{3H^{2}}{8\pi G}\left(
1+\frac{\dot{H}}{3H^{2}}\right)= \frac{3H^{2}}{8\pi G}\left(
1+\frac{aH^{'}}{3H}\right).
\end{equation}
Using $dt=da/aH$ it is easy to integrate Eq.(\ref{ff3})
\begin{equation}
\label{ppz} \phi=\int \left( -\frac{\dot{H}}{4\pi G}\right)^{1/2}
dt = \frac{1}{\sqrt{4\pi G}}\int
\left(-\frac{H^{'}}{aH}\right)^{1/2}da\;.
\end{equation}

Using Eq.(\ref{H(a)}) integration of (\ref{ff3})  in the interval
$[0,a]$ yields
\begin{eqnarray}\label{ppz1}\nonumber
\phi(a) & = & \frac{1}{\sqrt{2\pi G}n}\;{\rm sinh}^{-1}\left( \sqrt{D}a^{n}\right),\\
& = & \frac{1}{\sqrt{2\pi G}n}{\ln}\left(
\sqrt{D}a^{n}+\sqrt{Da^{2n}+1}\right)\;.
\end{eqnarray}

On its part, the potential energy takes the form
\begin{equation}
V(a)=\frac{H^{2}_{I}}{8\pi
G}\;\frac{3+Da^{2n}}{(1+Da^{2n})^{(n+2)/n}},
\end{equation}
or equivalently,
\begin{equation}
V(\phi)=\frac{H^{2}_{I}}{8\pi G}\; \frac{3+{\rm
sinh}^{2}(\sqrt{2\pi G} \;n \phi)} {[1+{\rm sinh}^{2}(\sqrt{2\pi
G}\; n \phi)]^{(n+2)/n}}  \;.
\end{equation}
Note that $V(0) =3H^{2}_{I}/8\pi G$ a value that should be
compared with the initial density as given by equation (14).
\bigskip



\begin{thebibliography}{99}
\bibitem{Transform} In order to obtain (12) we have applied the  transformation $u=-{1/D\,a^{2n}}$.
\end{thebibliography}
\end{document}